\documentclass{PoS}

\title{PHENIX plans for RHIC low energy run}

\ShortTitle{PHENIX plans for low energy}

\author{\speaker{Takao Sakaguchi}, for the PHENIX collaboration\\
        Physics Department, Brookhaven National Laboratory\\
        E-mail: \email{takao@bnl.gov}}

\abstract{
PHENIX plans for low energy running are presented. Current detector setting
makes it possible to measure dielectron spectra down to
$\sqrt{s_{NN}}$=39\,GeV, and photon/high $p_T$ hadron spectra down to
below sub-injection energy ($\sqrt{s_{NN}}$=5-10\,GeV). The upgrade of
the trigger scheme after the installation of VTX detector will enable
PHENIX to fully explore the sub-injection energy regime, starting 2011.
}

\FullConference{5th International Workshop on Critical Point and Onset of Deconfinement - CPOD 2009,\\
		 June 08 - 12 2009\\
		 Brookhaven National Laboratory, Long Island, New York, USA}

\begin{document}
\section{Introduction}
\begin{figure}[htbp]
\begin{minipage}{.65\textwidth}
\centering
\includegraphics[width=.9\textwidth]{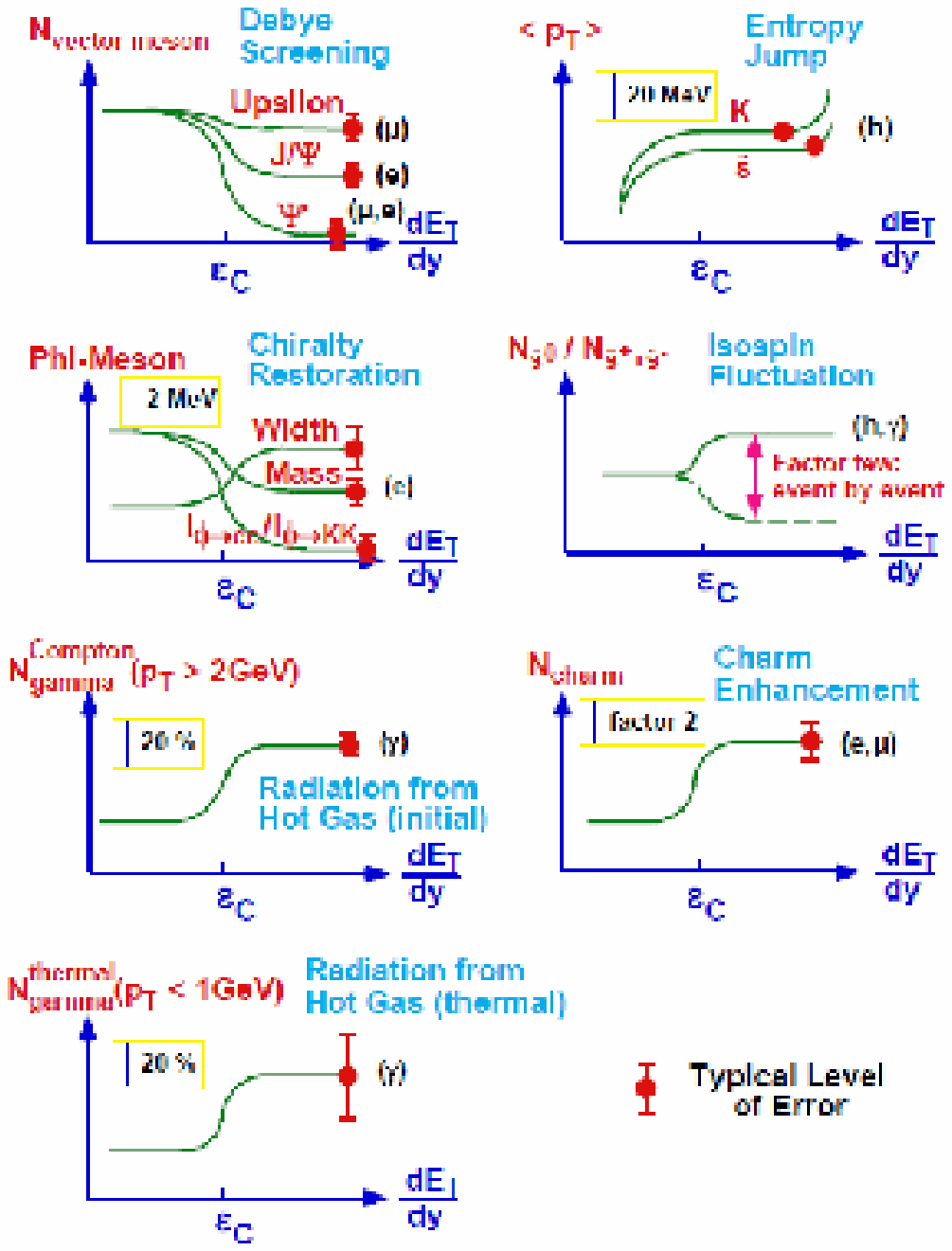}
\end{minipage}
\begin{minipage}{.33\textwidth}
\centering
\includegraphics[width=.9\textwidth]{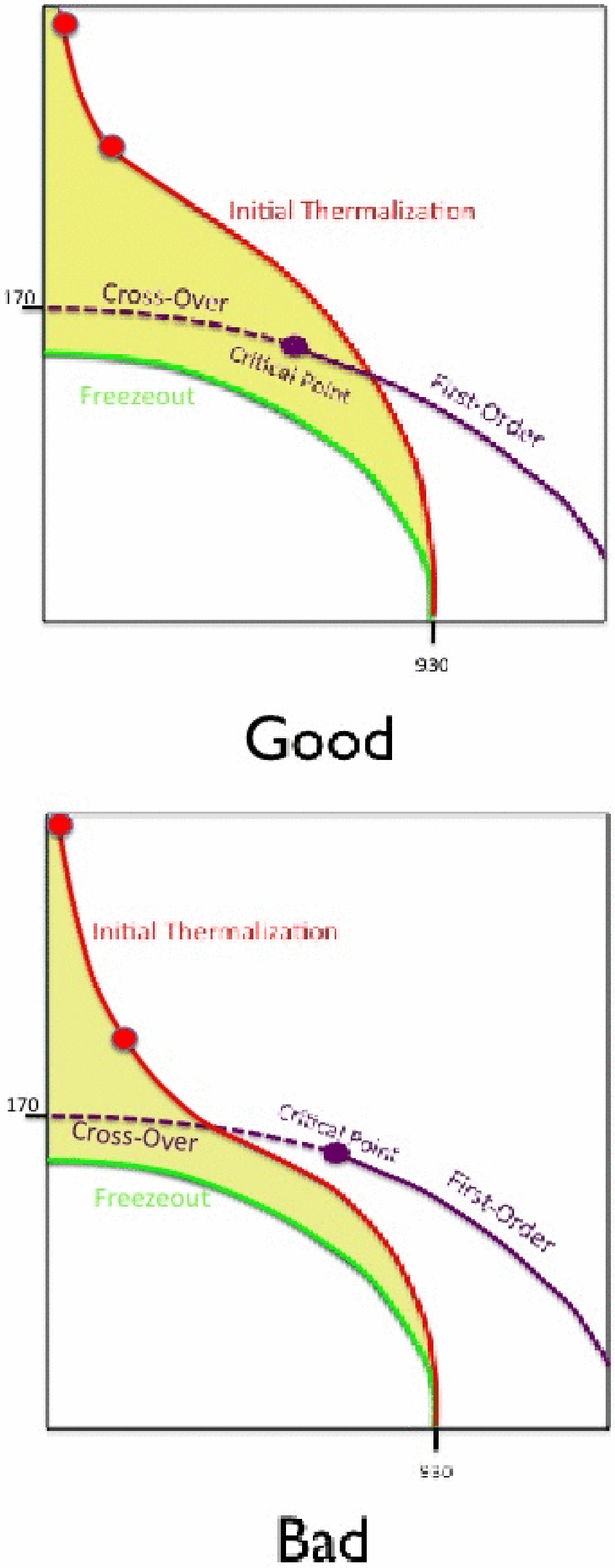}
\end{minipage}
\caption{Left: (a) expectation on crossing "the" QCD phase transition. Right: (b) Possible locations of critical points relative to freezeout and thermalization lines.}
\label{fig1}
\end{figure}

Since the start of RHIC running in 2000, many discoveries have been
made at RHIC not only in conventional observables that have been measured
at lower energies but also in new phenomena that were predicted but have never
been observed~\cite{ref1,ref2,ref3,ref4}. Figure~\ref{fig1}(a) shows the
cartoon of a prediction, drawn by Shoji Nagamiya, the first spokesperson
of the PHENIX collaboration, on what is expected for observables already
found in low $\sqrt{s}$ experiments, when we cross "the" boundary between
hadron and quark-gluon phases.
At RHIC, most of these observables have been measured, however, no critical
behavior has been seen in them so far. On the other hand, several new
phenomena have been discovered at RHIC, such as jet quenching or prompt
production, relying on hard scattering process, whose cross-section has
increased drastically at this high $\sqrt{s}$. High transverse momentum
($p_T$) particle correlations ($\gamma$-hadron and dihadron correlations)
that are also produced in the initial hard scattering were also measured at
RHIC for the first time. The RHIC results triggered a new look of SPS data,
and indeed many of these probes were found to have existed already at SPS
energies, however, these results were published only after the discovery
has been made at RHIC. This is an example how a high $\sqrt{s}$ machine
opened up a new field. It tells that the low energy running at RHIC is not
a repetition of exercises done in the past, but will study new phenomena
through new probes inaccessible so far due to insufficient statistics.
Therefore, the low energy running at RHIC is old in the sense of $\sqrt{s}$,
but is new in the sense that we are going to look at new probes.

There have been many studies on where the critical end point is located. The
question is whether or not this location is in between initial thermalization
and freezeout on QCD phase diagram. If not, no experiment may see the
signature of critical point (Fig.~\ref{fig1}(b)). Obviously, the search for the
critical point is not a problem dominated either by experiment or theory alone.
A close collaboration between experimentalists and theorists is essential.

\section{Issues in low energy running}
\begin{figure}[htbp]
\begin{minipage}{.5\textwidth}
\centering
\includegraphics[width=.9\textwidth]{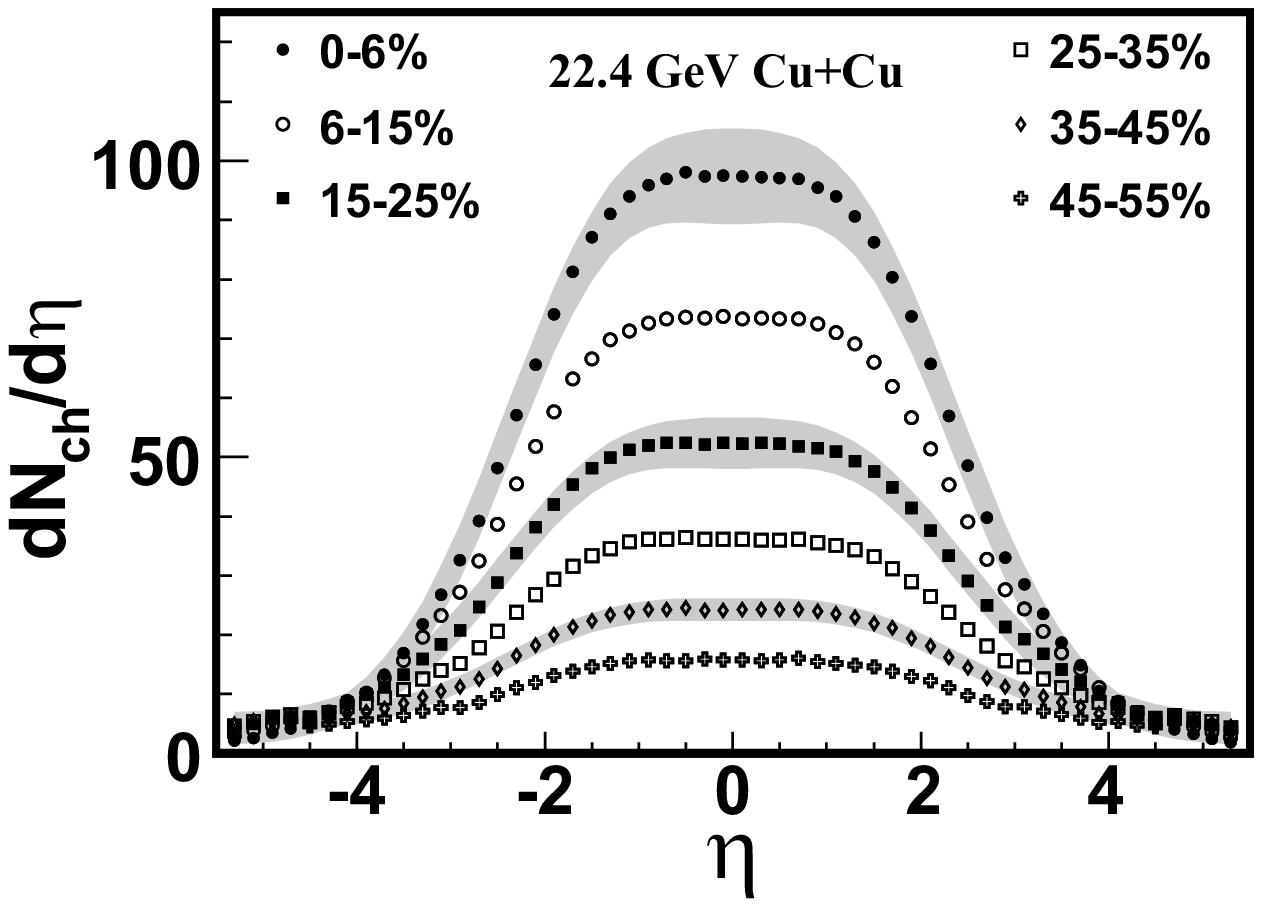}
\end{minipage}
\begin{minipage}{.5\textwidth}
\centering
\includegraphics[width=.9\textwidth]{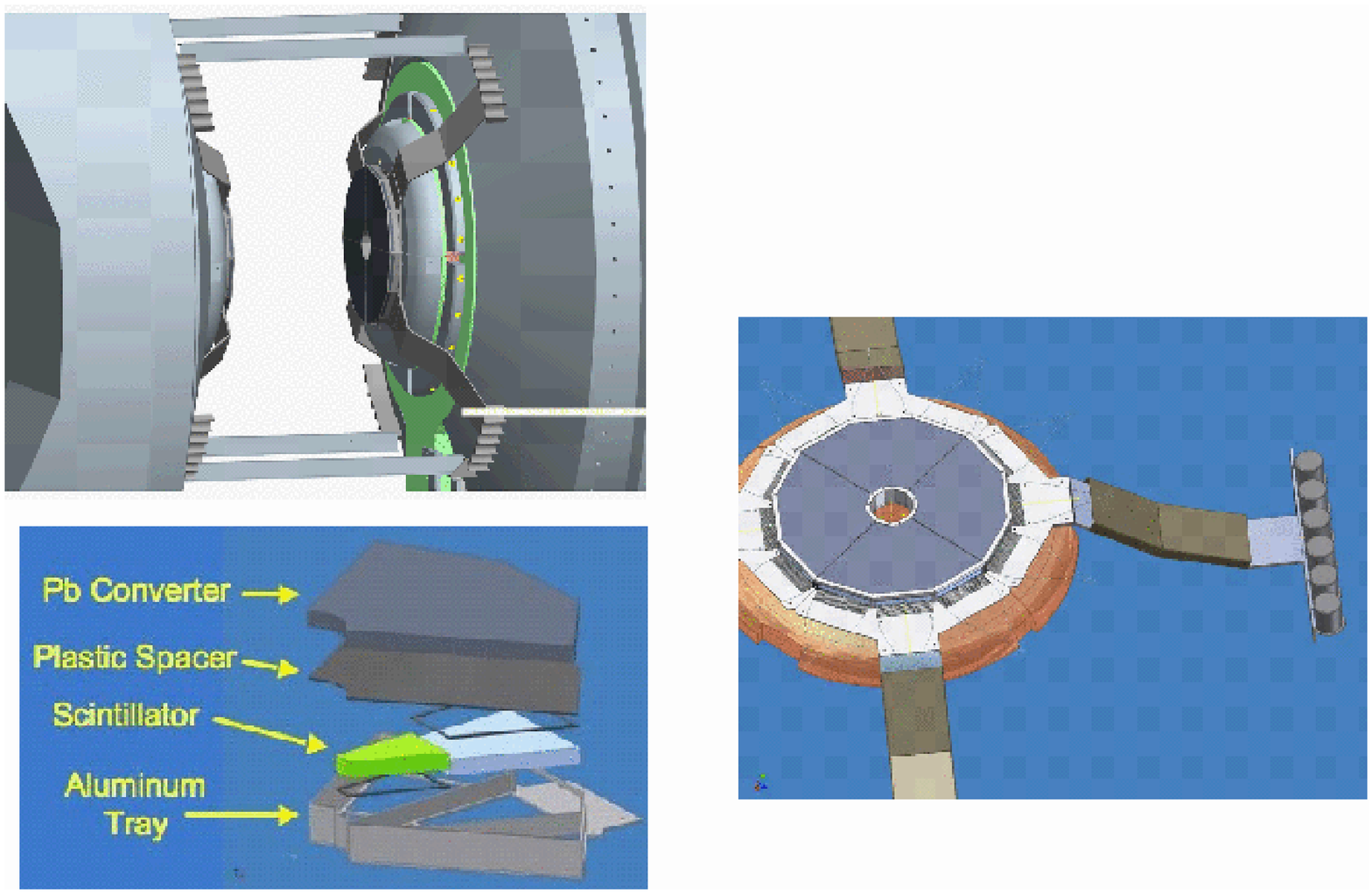}
\end{minipage}
\caption{Left: Charge multiplicity measurement at $\sqrt{s_{NN}}$=22.4\,GeV performed by PHOBOS experiment. Right: PHENIX RxNP detector.}
\label{fig3}
\end{figure}

Measurement at lower $\sqrt{s}$ should be relatively easy in terms of
multiplicity or particle identification (lower $\sqrt{s}$ system would not
produce as many and high energy particles as $\sqrt{s}$=200\,GeV did).
However, since the detector systems are optimized to measure these probes
and took some advantages from phenomena that are realized at higher
$\sqrt{s}$, running the detector at lower $\sqrt{s}$ may require addition of
a dedicated device. 
The most crucial part is triggering and characterizing events.
Because of very low multiplicity, it is possible that the trigger becomes
less efficient. Moreover the particles with high rapidity are dominantly
produced by fragmentation process, and thus the magnitude of the particle
multiplicity is no longer an excellent measure of collision centrality.
PHOBOS experiment at RHIC has measured pseudo-rapidity distributions in
Cu+Cu at various $\sqrt{s}$ (Fig.~\ref{fig3})~\cite{ref5}. As is seen from
the plot, at 3.1$<|\eta|<$3.9 where the beam-beam counter (BBC) of the
PHENIX detector is installed, the multiplicity is very low even at
$\sqrt{s}$=22.4\,GeV~\cite{ref6}.
The smaller number of detected particles will also result in degrading of
timing resolution, and thus the vertex position resolution becomes poorer.
PHENIX installed the reaction-plane (RxNP) detector that covers 1$<|\eta|<$2.8
before starting Run-7. It is for determining the reaction plane with a better
resolution than the one with the BBC. Since the multiplicity at this rapidity
region is higher than at 3.1$<|\eta|<$3.9 (BBC acceptance), it would give
a complementary information for triggering.

\begin{figure}[htbp]
\begin{minipage}{.4\textwidth}
\centering
\includegraphics[width=.9\textwidth]{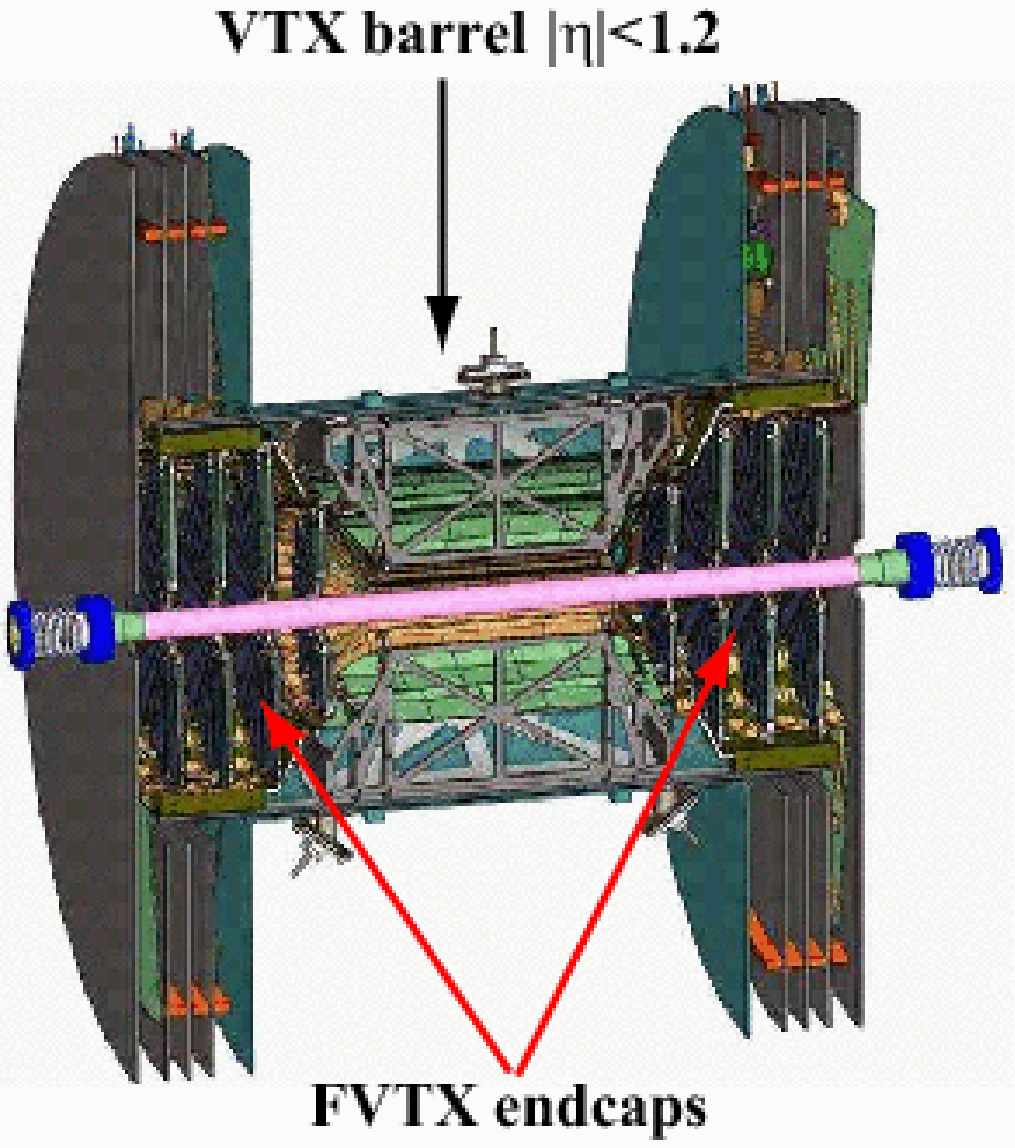}
\end{minipage}
\begin{minipage}{.6\textwidth}
\centering
\includegraphics[width=.9\textwidth]{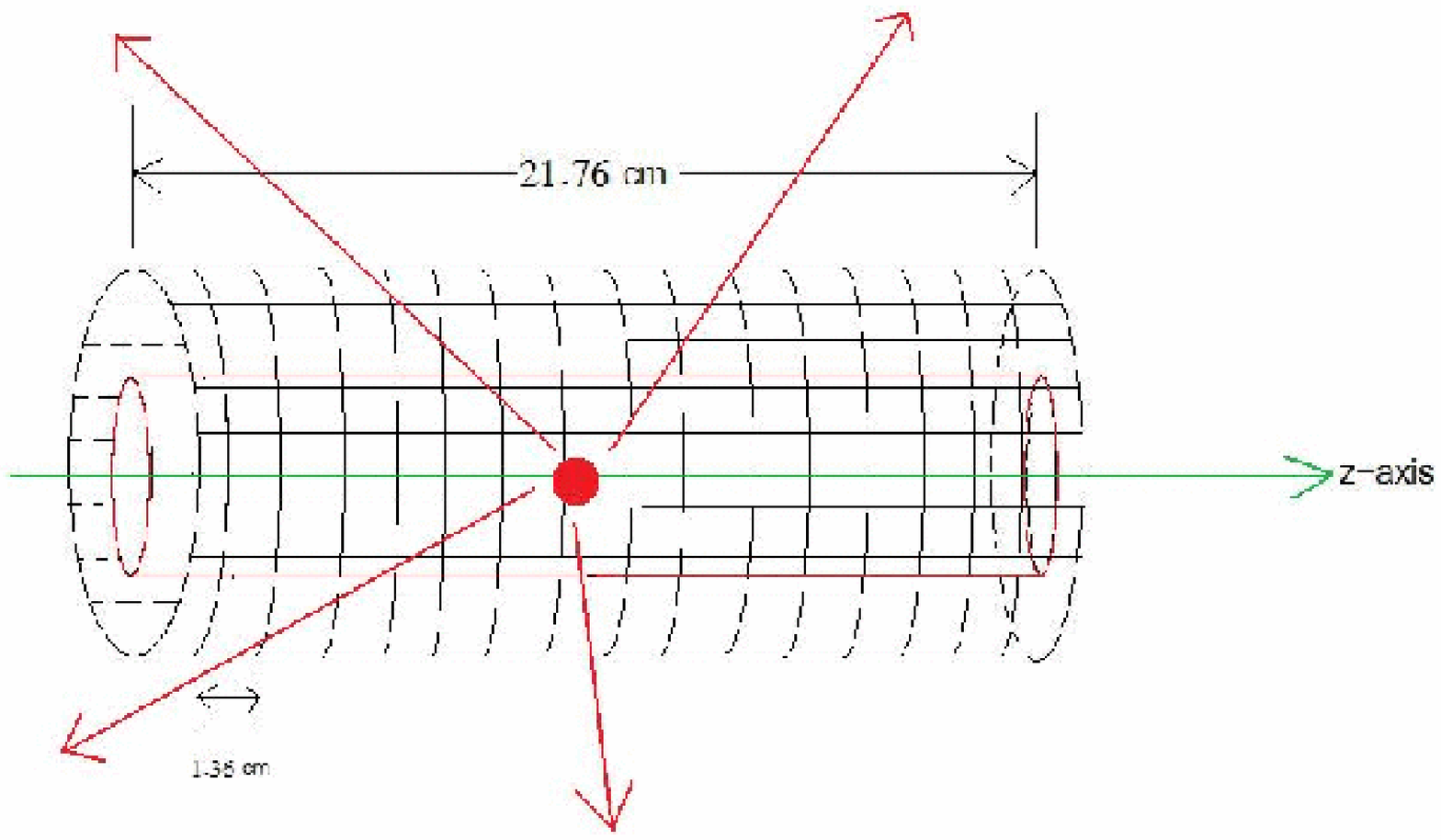}
\end{minipage}
\caption{Left: VTX detector. Right: trigger map using signals from VTX detector. The signals from each pad/strip will be summed and sent to global trigger system.}
\label{fig4}
\end{figure}

\begin{figure}[htbp]
\centering
\includegraphics[width=.8\textwidth]{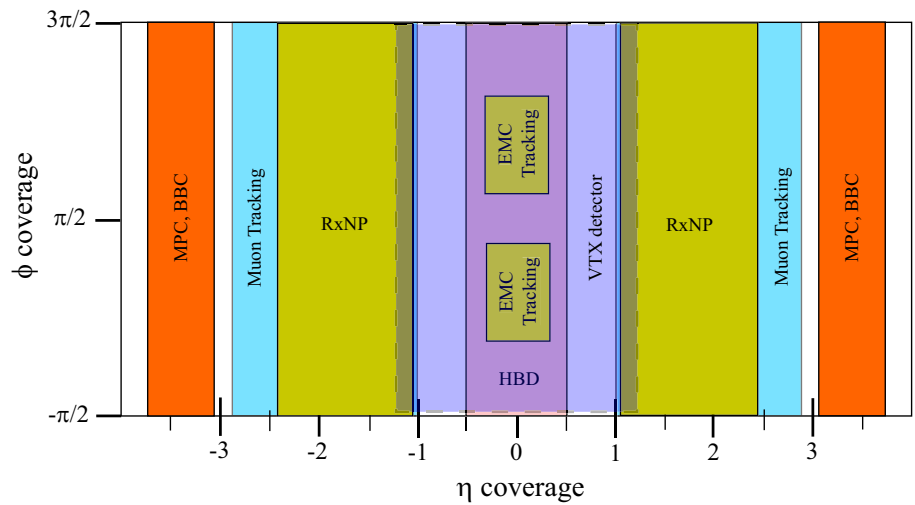}
\caption{Acceptance of the PHENIX detector after 2011. HBD is installed now, as shown in the schematics, but will be taken off in accordance with the installation of VTX detector.}
\label{fig4_2}
\end{figure}

Another issue at the low energies (around sub-injection energy) is rejection
of events not related to physics collisions, namely, beam-gas interactions,
beam scrape, and beam-beam interactions outside the vertex selections.
PHENIX is going to install a silicon vertex detector covering $|\eta|<$1.2
in 2011, and has a plan to make use of the signal from the detector to reject
such events at even lower energy such as sub-injection energy (5-10\,GeV)
(Fig.~\ref{fig4}).
As a result of these improvements, PHENIX will be fully efficient in
triggering and characterizing events at low $\sqrt{s}$. The expected
acceptance of the PHENIX detector by then is shown in Fig.~\ref{fig4_2}.
Before 2011, the VTX detector will not be installed because the same
space is occupied by a hadron blind detector (HBD)~\cite{ref7} which
targets a precise dielectron measurement in full energy Au+Au running in 2010.
The HBD can be operated in non-hadron blind mode, and positively detect
charged particles. However, it can not detect vertex points of the particles.

\section{Rich Physics with the PHENIX detector}
PHENIX can measure many fluctuation/HBT signals as well as identify charged
particles at low $p_T$ and neutral pions at high $p_T$. One of the
biggest advantage that PHENIX has is a large acceptance for leptons
and photons in the central arms that can scope rare leptonic and photonic
observables such as dielectron continuum and hadronic and non-hadronic
photons. Here we discuss the feasibility of measuring these observables
at lower $\sqrt{s}$.

\subsection{Nuclear modification factors of high $p_T$ hadrons}
\begin{figure}[htbp]
\begin{minipage}{.5\textwidth}
\centering
\includegraphics[width=.9\textwidth]{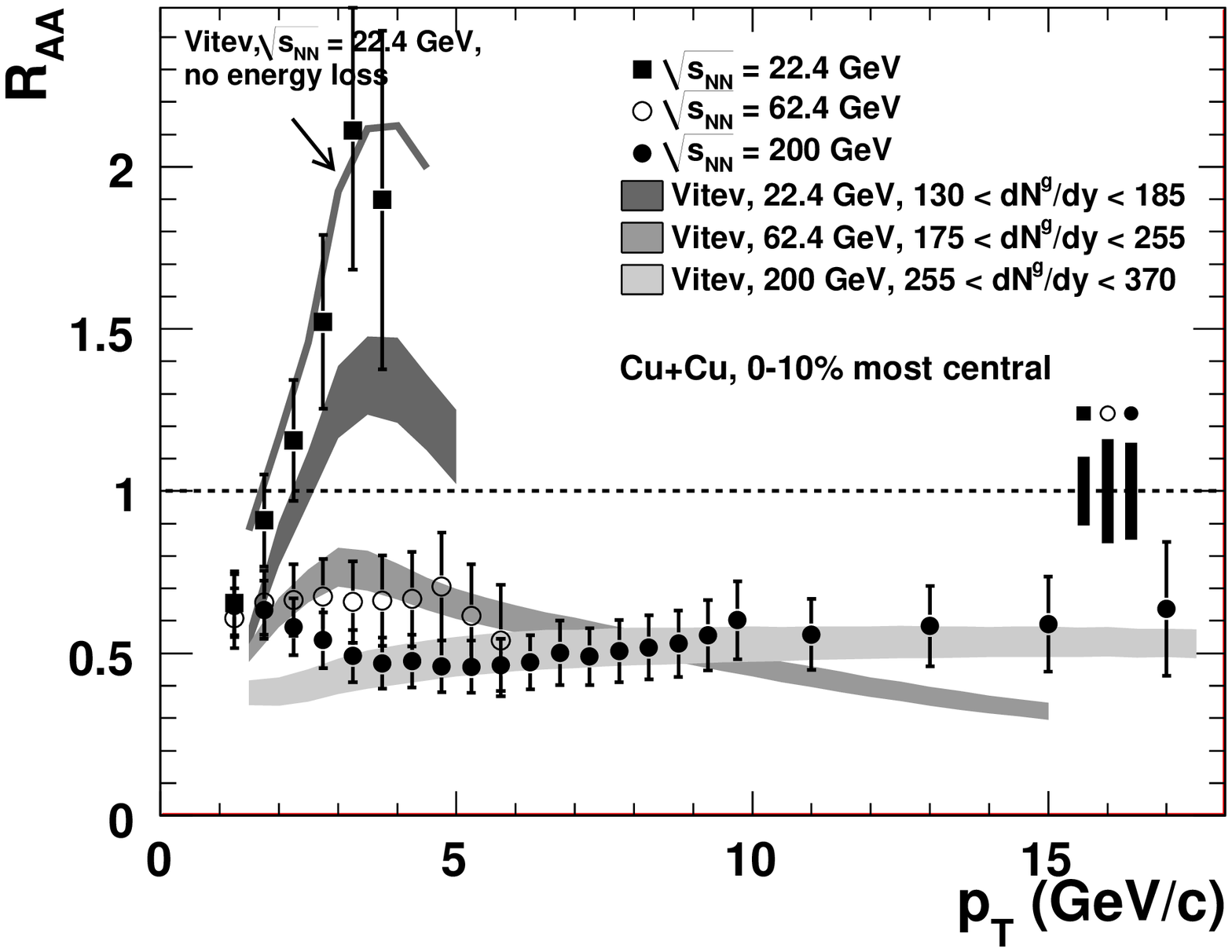}
\end{minipage}
\begin{minipage}{.5\textwidth}
\centering
\includegraphics[width=.9\textwidth]{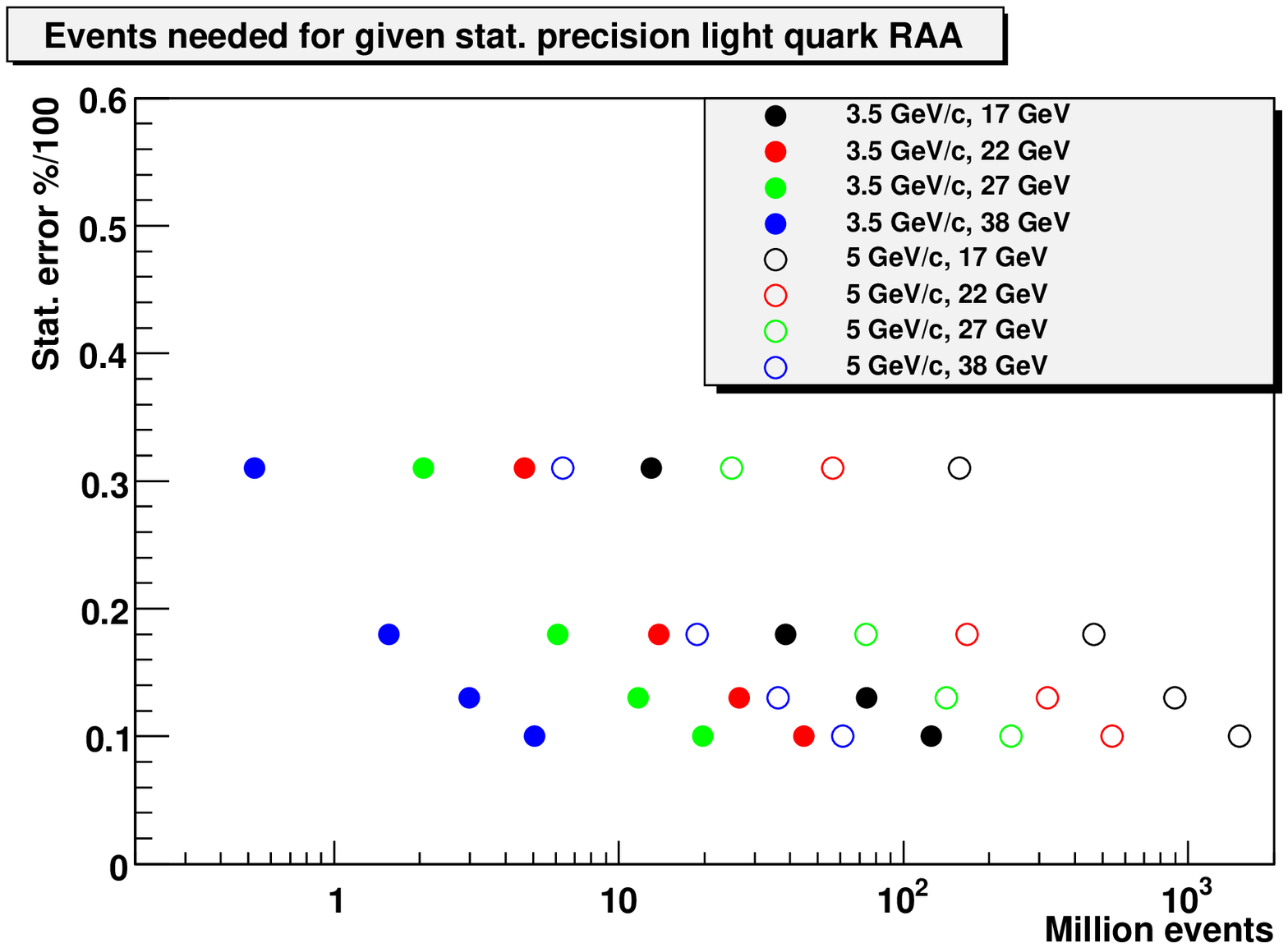}
\end{minipage}
\caption{Left: $\pi^0$ $R_{AA}$ in minimum bias Cu+Cu collisions as a function of $\sqrt{s_{NN}}$, performed by the PHENIX experiment. Right: significance of signal as a function of number of events recorded, at several $\sqrt{s_{NN}}$ points.}
\label{fig5_0}
\end{figure}

Nuclear modification factor ($R_{AA}$) of high $p_T$ hadrons (mainly
$\pi^0$ in our case) has been of
great interest in all the collision systems since we discovered the
suppression of the hadron yield in Au+Au collisions at
$\sqrt{s_{NN}}$=130\,GeV for the first time~\cite{ref8}, which was later
confirmed to be a consequence of an energy loss of hard scattered
partons by a control experiment in d+Au collisions~\cite{ref9} and
direct photon measurement in Au+Au collisions~\cite{ref10}.
The excitation function of $R_{AA}$ of high $p_T$ hadrons would be directly
connected to the parton (gluons or quarks) density of the
matter created. It is expected that in a low energy system, the partonic
degree of freedom vanishes and thus the energy loss of partons
disappears. At SPS energy ($\sqrt{s_{NN}}$=17.3\,GeV), it was understood
that the Cronin effect (initial multiple scattering of hard partons) dominates
the nucleus-nucleus collisions~\cite{ref11}. However, by taking a ratio of
A+A to p+A, it was found that there is a suppression of hadron yield even
at SPS~\cite{ref11_2}. The result from p+A collisions was assumed to be
a measure of the cold nuclear matter effect like Cronin, therefore the
A+A/p+A result indicated that the energy loss of hard partons had happened
even at this low energy. At RHIC, PHENIX has measured $R_{AA}$ of $\pi^0$'s
as a function of collision energies in Cu+Cu collisions as shown in
Fig.~\ref{fig5_0}~\cite{ref12}.
At 22.4\,GeV, the Cronin-like enhancement is seen, while at 62.4\,GeV and
200\,GeV, suppression is dominant. For 22.4\,GeV, unfortunately, p+p reference
data were not taken. Therefore, a complicated averaging of data taken by
other experiments had to be carried out to obtain the reference. In order to
fully accomplish 22.4\,GeV program, we need to measure hadron yields in
p+p collisions. It is one of the most urgent tasks in a low energy running.
In order to quantify the nuclear effect at 22.4\,GeV, we may want to take
d+A data as well. We estimated the number of events needed to measure
$R_{AA}$ with sufficient
significance for $\sqrt{s}$=17-38\,GeV, as shown in Fig.~\ref{fig5_0}.
We show the machine time needed to accumulate the events, later in this paper.

\subsection{Non-hadronic photon measurement}
\begin{figure}[htbp]
\begin{minipage}{.5\textwidth}
\centering
\includegraphics[width=.7\textwidth, angle=-90]{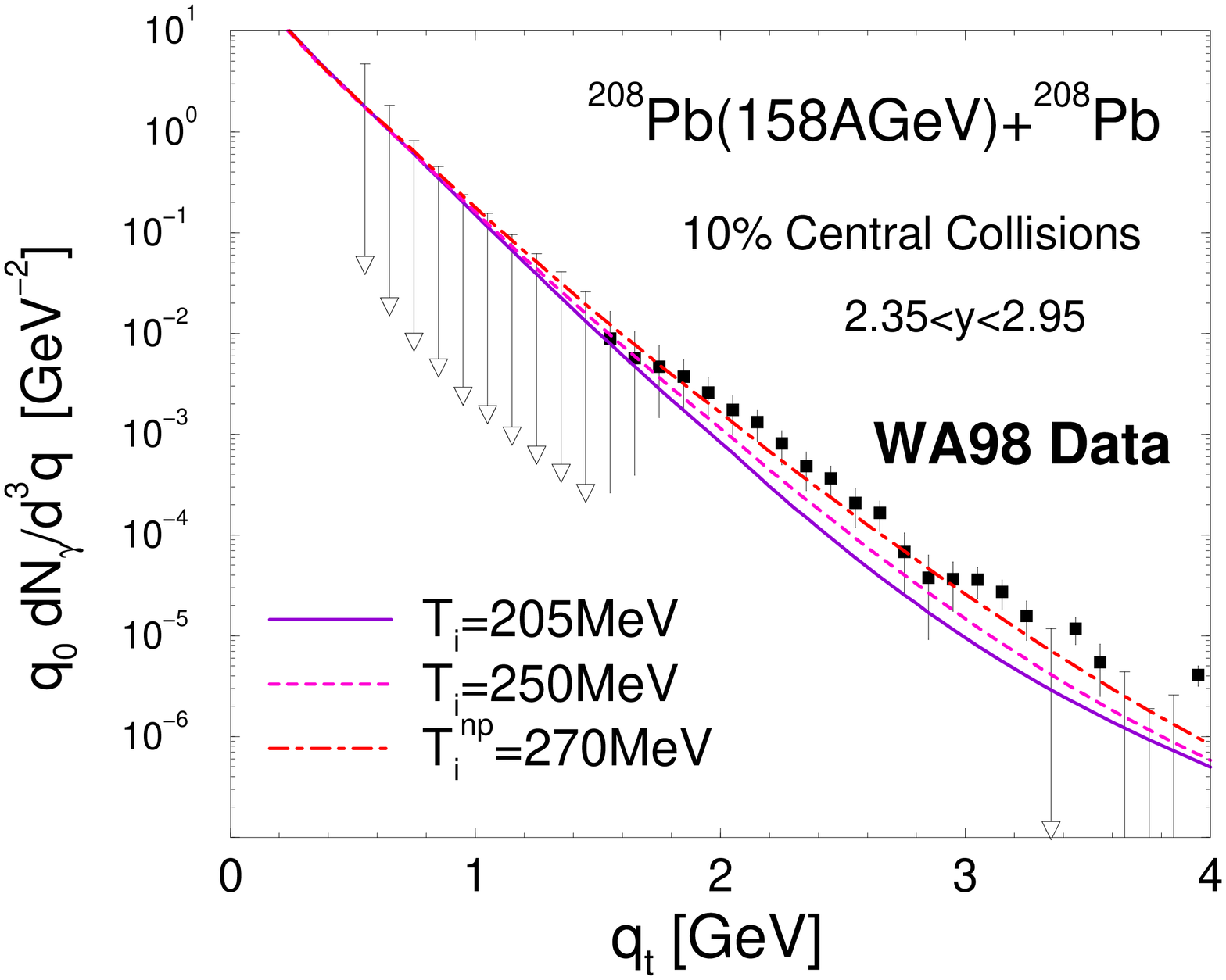}
\end{minipage}
\begin{minipage}{.5\textwidth}
\centering
\includegraphics[width=.7\textwidth, angle=-90]{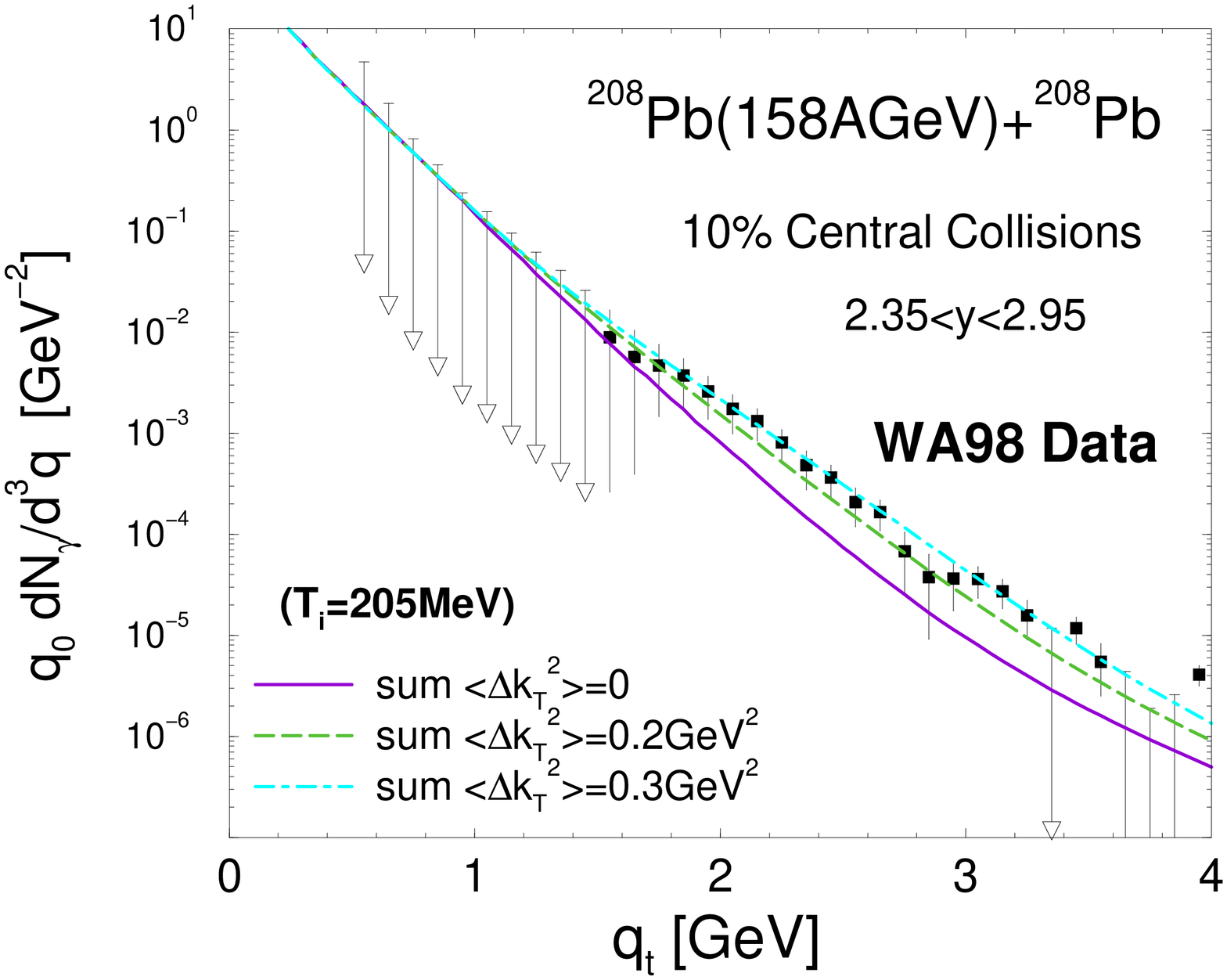}
\end{minipage}
\caption{WA98 results and comparison with theoretical calculation with various initial temperatures ($T_i$) and initial momentum broadenings ($<k_{T}>$).}
\label{fig5_1}
\end{figure}

Electro magnetic probes such as photons or leptons have a big advantage
over other observables in the sense that they are produced from all the
collisional stages and don't interact with medium once produced.
Therefore, by measuring the yields of the probes
as a function of $p_T$ and/or mass, we can obtain quantities related
to thermodynamical states of the collisional stages where photons or leptons
are produced. The first significant result on non-hadronic photons (photons
not from hadron decays) was obtained by the WA98 experiment~\cite{ref13}.
The result was analyzed by theorists to obtain the temperature of the system
produced. One of the many attempts is shown in Fig.~\ref{fig5_1}~\cite{ref14}.
The issue in analyzing the data was that there was no measurement of
hard direct photons (produced in initial hard scattering). Therefore,
the ratio of thermal photons and hard photons could not be determined,
and the data allowed many degrees of freedom in the theory calculation.
Had the $p_T$ range been extended to higher $p_T$ and confirmed the
contribution of hard photons, the initial temperature would have been
determined without large ambiguity. After years, WA98 has shown a preliminary
result of direct photons in p+Pb system~\cite{ref15}. The result does not
have significant signal, and therefore would not be a measure of initial
effect.

\begin{figure}[htbp]
\begin{minipage}{.5\textwidth}
\centering
\includegraphics[width=.9\textwidth]{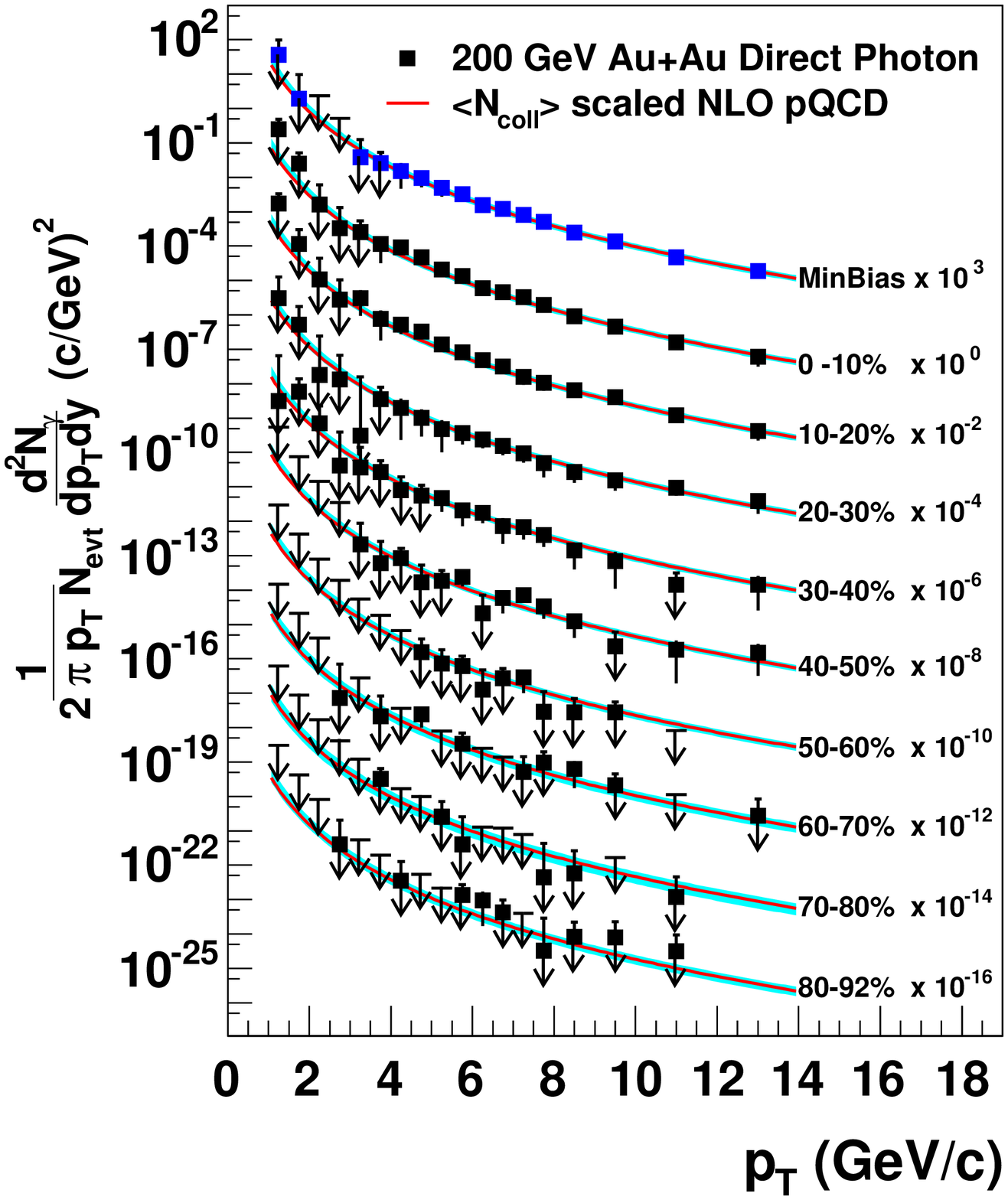}
\end{minipage}
\begin{minipage}{.5\textwidth}
\centering
\includegraphics[width=.9\textwidth]{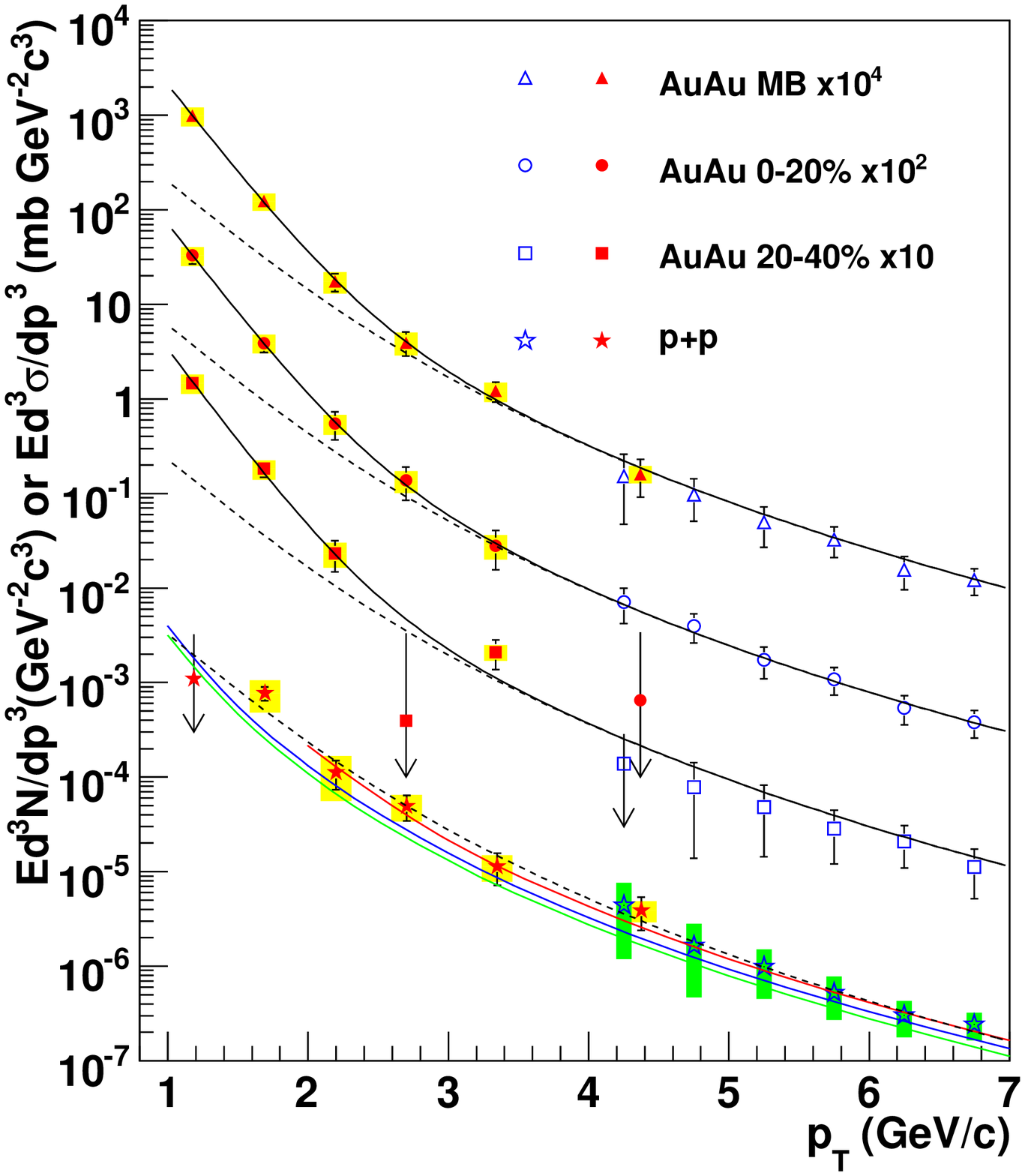}
\end{minipage}
\caption{High $p_T$ hard photon (left) and low $p_T$ photon (right) measurement by PHENIX.}
\label{fig5_2}
\end{figure}

The PHENIX experiment has measured hard direct photons in 2002 for the
first time in relativistic heavy ion collisions~\cite{ref10} as shown
in Fig.~\ref{fig5_2}.
It proved that the initial hard scattering was not suppressed and the
suppression of high $p_T$ hadrons was due to a final state effect.
The measurement of high $p_T$ photons is essential to establish the
baseline related to hard probes.

\begin{figure}[htbp]
\centering
\includegraphics[width=.6\textwidth]{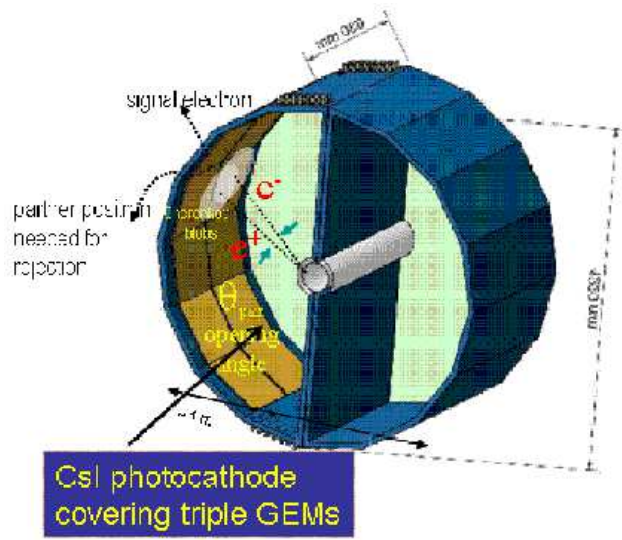}
\caption{Hadron blind detector (HBD) to tag Dalitz decay electrons from $\pi^0$ and conversion electrons.}
\label{fig6_0}
\end{figure}

For the low $p_T$, the signal is obscured by the large background of
$\pi^0$ and $\eta$ decays, whose yield is much higher than the expected
signal of thermal photons. A technique for dielectron measurement is applied,
assuming that the low mass high $p_T$ dielectrons are essentially the
virtual photons that are produced (although with much smaller probability)
in the same process of as that produce real photons. By selecting the mass
region of $m_{\pi}<m_{ee}<<p_T$, one can eliminate a large background from
$\pi^0$ Dalitz decays~\cite{ref16}. The dielectron yield is then converted
to real photon yield using Kroll-Wada formula~\cite{ref17}. The data obtained
with this method is shown in Fig.~\ref{fig5_2}. The signal to background
ratio will be much more improved by adding information from the hadron blind
detector (HBD), which was installed in PHENIX last year (Fig.~\ref{fig6_0}).

\subsection{Di-lepton measurement}
\begin{figure}[htbp]
\begin{minipage}{.5\textwidth}
\centering
\includegraphics[width=.9\textwidth]{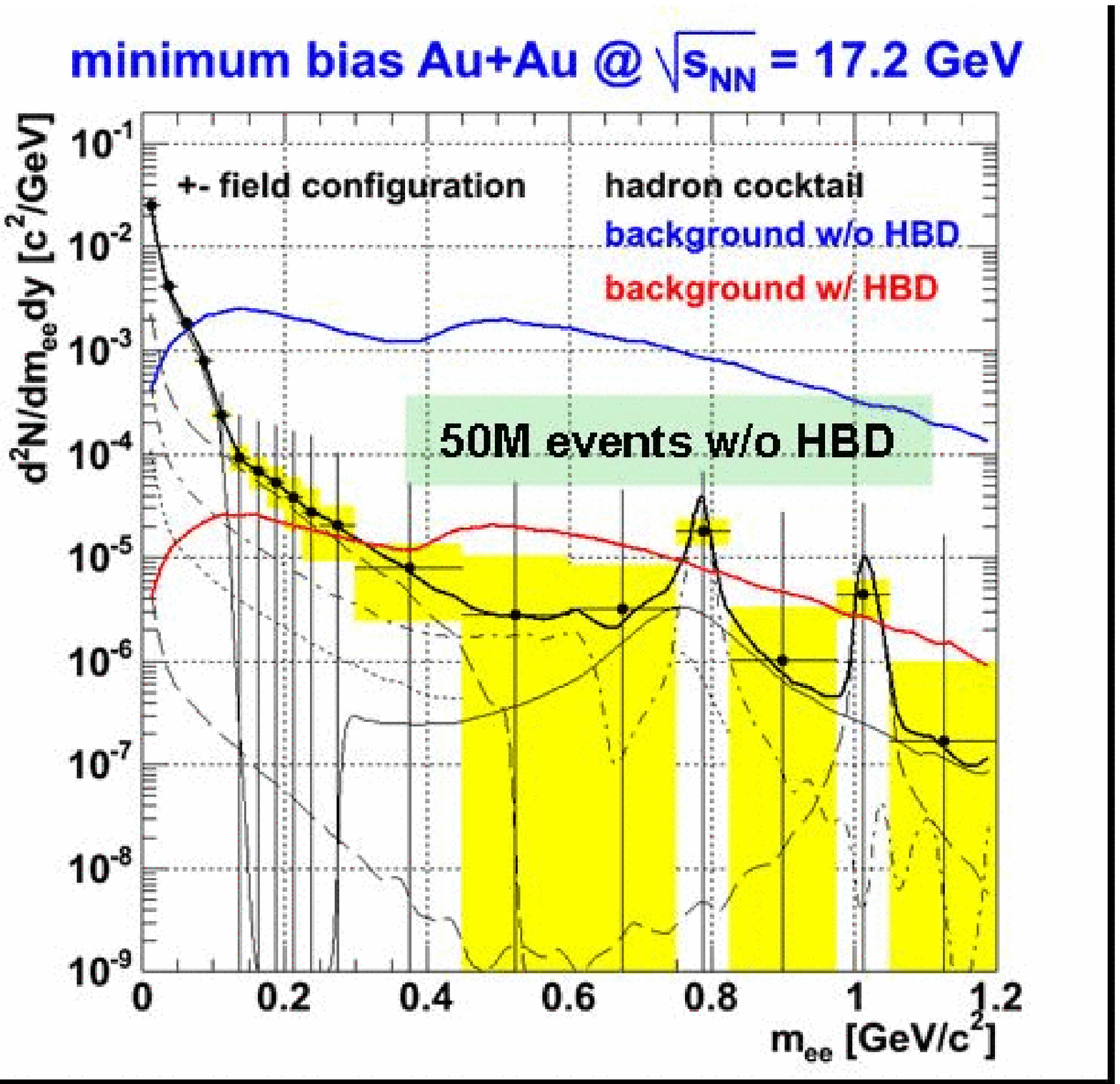}
\end{minipage}
\begin{minipage}{.5\textwidth}
\centering
\includegraphics[width=.9\textwidth]{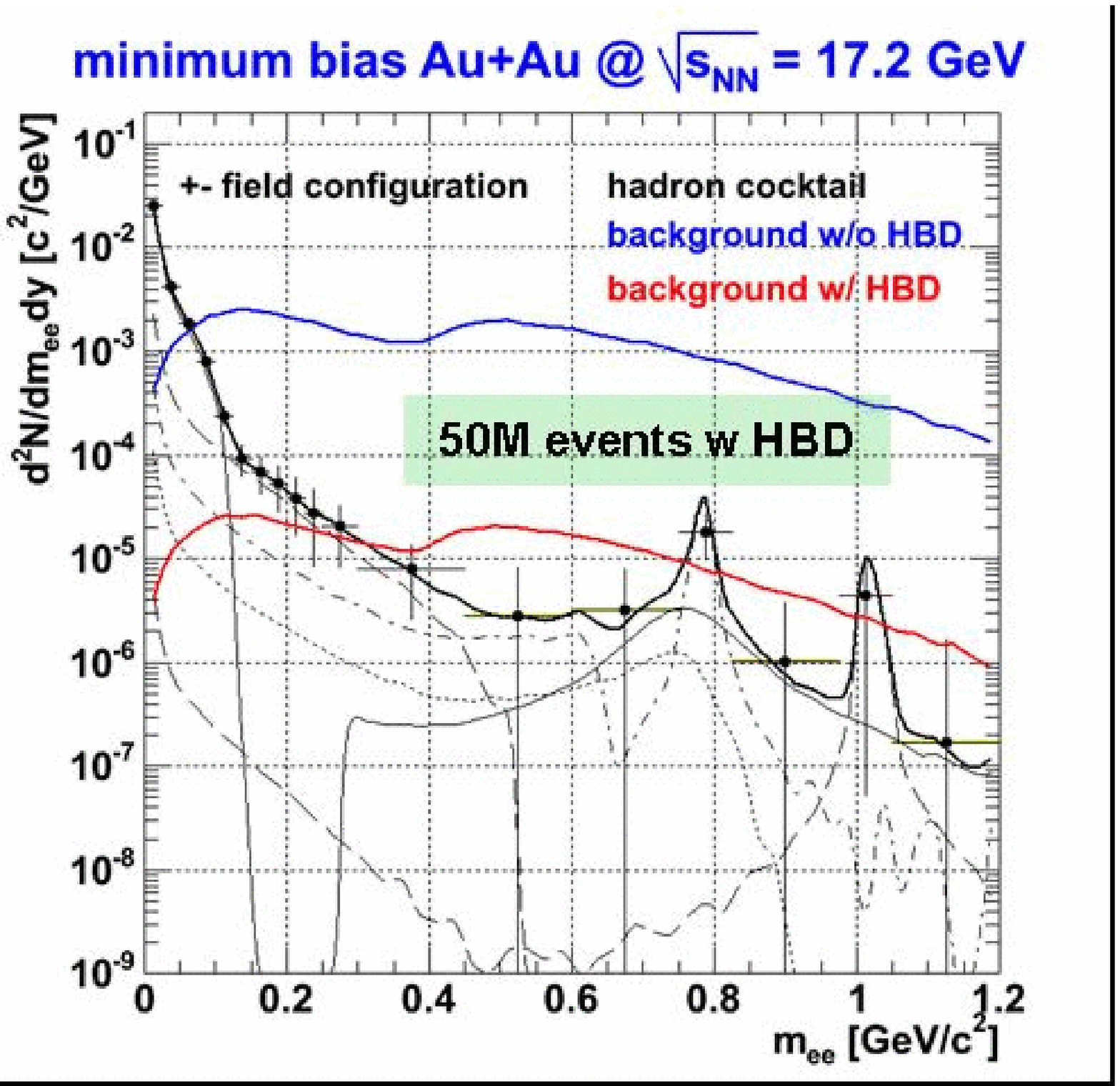}
\end{minipage}
\caption{Di-electron measurement without (left) and with (right) HBD at $\sqrt{s_{NN}}$=17\,GeV. The statistical errors are significantly reduced because the background Dalitz or photon conversion electrons are suppressed by the HBD.}
\label{fig7}
\end{figure}

Di-electrons have more degree of freedom in terms of slicing collisional
stages as well as efficiently eliminating backgrounds. The HBD detector in
PHENIX can tag Dalitz decay electrons from $\pi^0$ as well as electron pairs
produced via conversion of photons at beam pipe~\cite{ref7}. The HBD is a
Cherenkov detector consisting of CF$_{4}$ gas radiator, photo-cathodes and
three stage gas electron multipliers (GEMs), and will be operated in a field
free region. The electrons from $\pi^0$ Dalitz and photon conversions will
have small opening angle which will form a single cluster with higher charge
(corresponding to Cherenkov photo-electrons for two electrons), while single
electron will form a cluster with lower charge (i.e., Cherenkov
photo-electrons for one electron). The HBD detector will be kept installed
till the end of 2010, when a new detector, VTX detector, is going to be
installed to the same position to identify electrons from heavy flavor decays.
Therefore, these two years will be a golden opportunity for PHENIX to
measure dielectrons in heavy ion collisions. Fig.~\ref{fig7} shows the
possibility of measuring dielectrons with 50\,M events at
$\sqrt{s_{NN}}$=17\,GeV with and without HBD.
This shows the measurement in feasible as low as 17\,GeV. The significance
of the signal improves at higher energy. Before the trigger upgrades, the
PHENIX detector is less efficient in triggering at sub-injection energies,
but with the HBD, has a big advantage of measuring such rare probes as
dielectrons. Therefore, PHENIX may want to have a beam at higher energies
next year.

\section{Event rates}
The estimated number of minimum bias Au+Au collisions to make a significant
measurement with the PHENIX detector for each of the observables considered
is summarized in Table~\ref{tab:Summary-0}. 

\begin{table}
{\small
\begin{tabular}{|c|c|c|c|cc|cccc|c|c|cc|c|ccc|}
\hline
\rotatebox{90}{$\sqrt{s_{NN}}$}
& \rotatebox{90}{Fluctuations in $\langle n \rangle$}
& \rotatebox{90}{Fluctuations in $\langle p_t \rangle$}
& \rotatebox{90}{PID spectra, identified particle ratios}
& \rotatebox{90}{longitudinal density correlations} 
& \rotatebox{90}{critical exponent $\eta$} 
& \rotatebox{90}{1D imaging of pion source} 
& \rotatebox{90}{L\'evy exponent $\alpha$} 
& \rotatebox{90}{3d Gaussian HBT radii for pions $R_i(m_T)$} 
& \rotatebox{90}{HBT intercept parameter  $\lambda(m_T)$} 
& \rotatebox{90}{dielectron spectra, fluctuations in $\langle K/\pi\rangle $} 
& \rotatebox{90}{dihadron correlations} 
& \rotatebox{90}{nuclear modification factor $R_{AA}$} 
& \rotatebox{90}{optical opacity $\kappa$ } 
& \rotatebox{90}{heavy flavor electrons} 
& \rotatebox{90}{1d image of kaon source} 
& \rotatebox{90}{3d image of pion source} 
& \rotatebox{90}{Kaon 3d Gaussian HBT radii $R_i(m_T)$} 
\\
\hline
5.5  & 
0.01&	0.03&	0.03 &\multicolumn{2}{c|}{2}&\multicolumn{4}{c|}{54}& 50 & 375& \multicolumn{2}{c|}{NA}& NA & \multicolumn{3}{c|}{953}
 \\ \hline
7.7  & 
0.01&	0.03&	0.02 &\multicolumn{2}{c|}{2}&\multicolumn{4}{c|}{33}& 50 & 246& \multicolumn{2}{c|}{NA}& NA & \multicolumn{3}{c|}{586}
 \\ \hline
11.5 & 
0.01&	0.03&	0.02 &\multicolumn{2}{c|}{2}&\multicolumn{4}{c|}{24}& 50 & 160& \multicolumn{2}{c|}{NA}& NA & \multicolumn{3}{c|}{431}
 \\ \hline
17.3 & 
0.01&	0.03&	0.01 &\multicolumn{2}{c|}{2}&\multicolumn{4}{c|}{19}& 50 & 109& \multicolumn{2}{c|}{157}& NA & \multicolumn{3}{c|}{340}
 \\ \hline
27   & 
0.01&	0.03&	0.01 &\multicolumn{2}{c|}{2}&\multicolumn{4}{c|}{16}& 50 &  68& \multicolumn{2}{c|}{ 24}& NA & \multicolumn{3}{c|}{276}
\\ \hline
39   & 
0.01&	0.03&	0.01 &\multicolumn{2}{c|}{2}&\multicolumn{4}{c|}{14}& 50 &  48& \multicolumn{2}{c|}{6.3}& 700 & \multicolumn{3}{c|}{239}
 \\ \hline
\end{tabular}
}
\caption{\label{tab:Summary-0} 
Number of million Au+Au events needed to make a significant 
measurement with the PHENIX detector at various energies in the 
RHIC Low Energy Scan program.}
\end{table}
A rate estimate was performed based upon the rates and efficiencies observed
in the 9\,GeV Au+Au test run and the corresponding 19.6\,GeV and 62.4\,GeV
data. Peak luminosity projections for 2011 were assumed. Below injection
energy an improved luminosity of a factor of 6.0 was assumed~\cite{ref18}.
The main source of this improvement comes from filling all available bunches.
In the Run-9 9.2\,GeV Au+Au data analysis, a significant fraction of bad
events was found, likely from beam-gas collisions and interactions with the
beam pipe. Simply requiring a reconstructed BBC z-vertex was not sufficient
to isolate good collisions. The additional requirement of a reconstructed
ZDC z-vertex is necessary to obtain a clean sample. However, at such low
energies the fraction of the neutral particles that generates hits in the
ZDC reduces drastically. Correspondingly, the required inclusion of the ZDC
in the event cuts is found to be  the primary reason that only 3\% of the
BBCLL1 triggered events were available for analysis. Any improvement in the
PHENIX triggering capabilities that does not require a coincidence with the
ZDC would greatly improve the event purity at low energies and thus greatly
decrease the amount of time necessary to achieve sufficient statistics
for the observables of interest. This is demonstrated by
Table \ref{table:Summ-1} which show running time estimates for best and
worst cases.
\begin{table}
\begin{center}
\begin{tabular}{cccc}
$\sqrt{s_{NN}}$ & Minimum 10-hour-days & Maximum 10-hour-days\\ \hline
5.0 & 14.3 & 3714 \\
7.7 & 6.9 &  623.4 \\
11.5 & 3.5 & 119.0 \\
17.3 & 1.8 & 22.0 \\
27.0 & 0.4 & 2.5 \\
39.0 & 0.1 & 0.4 \\
\end{tabular}
\end{center}
\caption{\label{table:Summ-1}
Time estimates for accumulating 1M events based only upon the rates and
efficiencies observed in the 9\,GeV Au+Au test run with an improved
luminosity of a factor of 6.0 (Maximum). Time estimates based upon the
rates the assumption that the luminosity can be improved by a factor
of 1.8 by 2012 and that the luminosity below injection energy improves
by a factor of 6 along with the assumption that the recorded data purity
is 100\,\% (Minimum).
}
\end{table}
In this estimate, we used HIJING as an event generator, which does not
include Fermi motion effect. A preliminary study using UrQMD shows that
the trigger efficiency at $\sqrt{s_{NN}}<$10\,GeV is much higher than
the one we found using HIJING. A further detailed study is ongoing.

\section{Summary}
PHENIX plans for low energy running are presented. Current detector setting
makes it possible to measure dielectron spectra down to
$\sqrt{s_{NN}}$=39\,GeV, and photon/high $p_T$ hadron spectra down to
below sub-injection energy ($\sqrt{s_{NN}}$=5-10\,GeV). A further study of
trigger efficiency is ongoing using the UrQMD event generator. The upgrade of
the trigger scheme after the installation of VTX detector will enable
PHENIX to fully explore the sub-injection energy regime, starting 2011.

\end{document}